\documentclass{nature}
\usepackage[T1]{fontenc}
\usepackage{lmodern}
\usepackage{cuted}

\usepackage[fixed]{fontawesome5}

\usepackage[defaultsans]{comfortaa}
\usepackage{inconsolata}   
\usepackage{amsfonts}
\usepackage{bm}
\usepackage{float}
\usepackage{physics}
\usepackage{upgreek}
\usepackage{caption}
\usepackage{graphicx}
\usepackage{xcolor}
\definecolor{lightgray}{gray}{0.93} 
\usepackage[top=1.27cm, bottom=2.5cm, left=1.27cm, right=1.27cm]{geometry}
\usepackage{multicol}
\usepackage{hyperref}
\hypersetup{
    colorlinks=true,
    linkcolor=blue,
    filecolor=blue,
    urlcolor=blue,
    citecolor=blue,
}

\usepackage{soul}     
\usepackage{lipsum}   
\usepackage{fancyhdr}

\makeatletter
\let\saved@includegraphics\includegraphics
\AtBeginDocument{\let\includegraphics\saved@includegraphics}

\makeatother
\title{Quantum Light Detection with Enhanced Photonic Neural Network}
\author{S.~Świerczewski$^{\text{1,2}}$, D. Ko$^\text{3}$, A. Rahmani$^\text{3}$, J. C. López Carreño$^\text{{1,2}}$,  W. Verstraelen$^\text{{4,5}}$,
P. Deuar$^\text{{3}}$, B.~Pi\k{e}tka$^\text{{1}}$, T. C. H. Liew$^\text{{4,5}}$, M.~Matuszewski$^\text{{2,3}}$, A.~Opala$^\text{{1,2,3,\scriptsize{\faMicrochip}}}$}
\begin{document}
\begin{spacing}{1}
\maketitle
\thispagestyle{empty}
\vspace{2mm}
\begin{scriptsize}
\begin{affiliations}
\item Institute of Experimental Physics, Faculty of Physics, University of Warsaw, ul. Pasteura 5, PL-02-093` Warsaw, Poland
\noindent $^{2}$ Center for Theoretical Physics, Polish Academy of Sciences Aleja Lotników 32/46, 02-668 Warsaw, Poland
\noindent $^{3}$ Institute of Physics, Polish Academy of Sciences, Aleja Lotnik\'ow 32/46, PL-02-668 Warsaw, Poland
\noindent $^{4}$ Division of Physics and Applied Physics, School of Physical and Mathematical Sciences, Nanyang Technological University, Singapore, Singapore, \noindent $^{5}$ Majulab, International Joint Research Unit UMI 3654, CNRS, Université Côte d’Azur, Sorbonne Université, National University of Singapore, Nanyang Technological University, Singapore, Singapore 117543,
${\faMicrochip}$- corresponding author: aopala@fuw.edu.pl
\end{affiliations}
\end{scriptsize}

\begin{abstract}
\noindent
\colorbox{lightgray}{%
  \parbox{\dimexpr\linewidth-2\fboxsep}{%
  \scriptsize
  \vspace{2mm}
  {Advances in quantum technologies are accelerating the demand for optical quantum state sensors that combine high precision, versatility, and scalability within a unified hardware platform. Quantum reservoir computing offers a powerful route toward this goal by exploiting the nonlinear dynamics of quantum systems to process and interpret quantum information efficiently. Photonic neural networks are particularly well suited for such implementations, owing to their intrinsic sensitivity to photon-encoded quantum information. However, the practical realisation of photonic quantum reservoirs remains constrained by the inherently weak optical nonlinearities of available materials and the technological challenges of fabricating densely coupled quantum networks. To address these limitations, we introduce a hybrid quantum–classical detection protocol that integrates the advantages of quantum reservoirs with the adaptive learning capabilities of analogue neural networks. This synergistic architecture substantially enhances information-extraction accuracy and robustness, enabling low-cost performance improvements of quantum light sensors. Based on the proposed approach, we achieved significant improvements in quantum state classification, tomography, and feature regression, even for reservoirs with a relatively small nonlinearity-to-losses ratio $U/\gamma \approx 0.02$ in a network of only five nodes. By reducing reliance on material nonlinearity and reservoir size, the proposed approach facilitates the practical deployment of high-fidelity photonic quantum sensors on existing integrated platforms, paving the way toward chip-scale quantum processors and photonic sensing technologies.}}}
\end{abstract}
 \vspace{0.5mm}
\begin{multicols}{2}
\scriptsize

\section*{\normalsize Introduction}

Quantum sensing has redefined the limits of precision measurement, achieving sensitivities that surpass classical systems. By exploiting squeezed and entangled states, state-of-the-art quantum devices, including atomic clocks~\cite{Finkelstein_2024}, magnetometers~\cite{Casola_2018}, and interferometers~\cite{Pirandola_2018, Jia_2024}, are advancing both fundamental science and real-world applications.

As quantum technology advances toward scalable, error-resilient applications, there is an increasing need for high-fidelity quantum-state sensors that operate in real time. In the near future, quantum sensors are expected to play a central role in quantum metrology~\cite{Bongs_2023}, calibration, validation, or control of next-generation quantum devices across fields of quantum information processing~\cite{Cerezo_2021} and quantum machine learning~\cite{Biamonte_2017, Cong_2019}. 

In parallel, neuromorphic computing (NC) is reshaping paradigms of modern information processing by introducing brain-inspired hardware that overcomes the von Neumann bottleneck~\cite{Marković2020_2, Schuman_2022, Kudithipudi_2025}. 

Among emerging NC platforms, integrated nano- and micro-photonic systems stand out, enabling femtojoule-level multiply–accumulate operations and terahertz-scale throughput~\cite{Shastri_2021, Ma_2025, Matuszewski_2024, Farmakidis_2024}. Their single-photon sensitivity~\cite{Roberta_2024, Maring_2024}, programmability, and on-chip integration~\cite{Arrazola_2021, Maring_2024} have positioned photonic architectures as a natural platform for cutting-edge classical~\cite{Wetzstein_2020} and quantum optical neural network (QONN) implementations. 

QONN systems combine brain-inspired information processing~\cite{Ghosh_2019, Marković_2020, Ghosh_2021} with the capabilities of quantum photonic hardware. Recent advances in integrated photonics~\cite{Dusanowski_2025, Madigawa_2025}, high-Q topological cavities, memristive devices~\cite{Spagnolo_2022}, quantum dots~\cite{Dusanowski_2019, Uppu_2021, Dusanowski_2023, Wang_2019, Huang_2025, Yao_2025}, and exciton–polariton microcavities~\cite{Fink_2018, Sanvitto_2018, Klembt_2021, Gianfrate_2024} already set solid foundation for the next generation of quantum technologies~\cite{Pelucchi_2022, Wang_2025, Couteau_2023} with high potential for QONNs realization.

Within this framework, quantum reservoir computing (QRC) has emerged as a powerful paradigm for designing QONN systems. By exploiting the dynamical response of nonlinear quantum systems to probe quantum states, QRC enables state classification, tomography, and circuit emulation~\cite{Ghosh_2019, Krisnanda_2022, Spagnolo_2022, Krisnanda_2023TM, Xu_2023, Ko_2025}. Relying on intrinsic nonlinear dynamics rather than precise internode control, QRC offers an experimentally practical framework, making it highly attractive for photonic hardware implementations~\cite{Spagnolo_2022}.

However, bosonic photonic quantum reservoirs face a fundamental limitation arising from intrinsically weak optical nonlinearities. Unlike fermionic systems, where Pauli exclusion enforces strong interactions, the nonlinear character of bosonic systems depend on engineered Kerr-like, cross-Kerr, or parametric processes~\cite{Xu_2023, DiCandia2023, Kounalakis_2018}, which remain extremely challenging to realize at the single-photon level in scalable devices. Such a constraint limits the performance of bosonic QRC and hinders its deployment for large-scale quantum photonic machine learning and intelligent quantum sensing~\cite{Ma_2022}.

Here, we introduce a hybrid quantum–classical reservoir architecture for quantum light sensing that overcomes this bottleneck. By coupling a bosonic reservoir with an external hardware or software neural network, the proposed approach compensates for the weak nonlinearities of photonic systems, enabling advanced quantum machine-learning tasks. Through theoretical analysis, we demonstrate and quantify a significant enhancement in QRC-based sensor performance for quantum-state classification, tomography, and feature-prediction tasks, achieving error reductions previously unattainable in bosonic systems operating in weak or intermediate nonlinear regimes. Importantly, the proposed method is fully compatible with existing integrated photonic platforms, requiring neither additional reservoir engineering nor explicit quantum circuit programming.


\section*{\normalsize Results}

\begin{figure*}[bht!]
    \centering
    \includegraphics[width=0.830\linewidth]{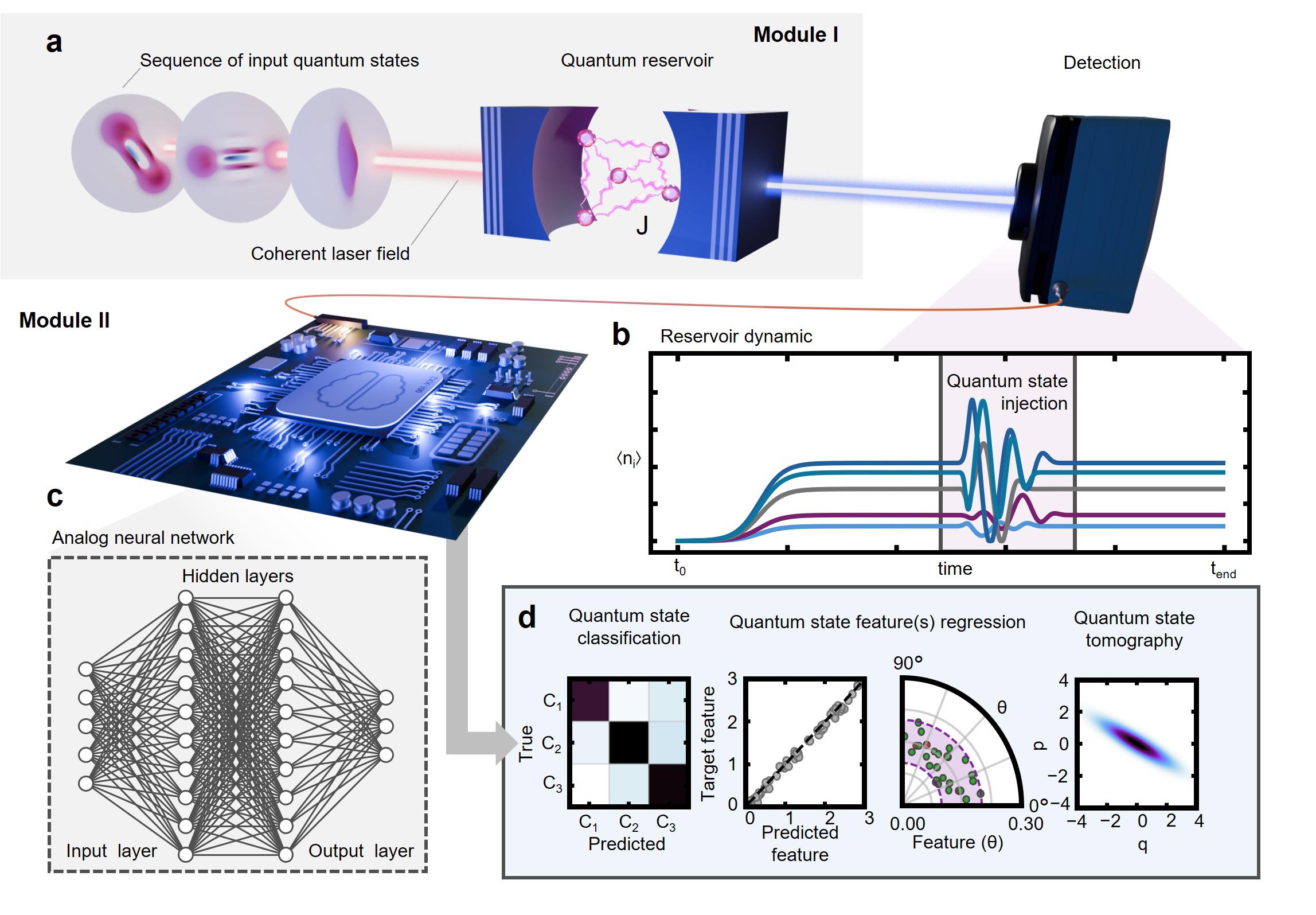}
    \caption{\scriptsize\textbf{\textbf{Enhanced photonic reservoir neural network for multi-task quantum state sensing.}}
\textbf{a} The sequence of probing quantum states and coherent laser field acting on the optical resonator. An input quantum state drives the dynamics of an optical reservoir formed by coupled bosonic modes within a cavity, sequentially. 
\textbf{b} Time-dependent occupations of the reservoir nodes, encoding information about the input state’s density matrix, are continuously monitored by a detection system (schematic illustration). 
\textbf{c} The recorded reservoir responses are processed by a classical feed-forward neural network, implemented either in software or hardware, which is trained to extract task-relevant quantum-state features from nonlinear reservoir dynamics. 
\textbf{d} The hybrid sensor performs multiple quantum sensing tasks, including state classification, parameter regression, and quantum state tomography.}
    \label{fig:fig1}
\end{figure*}

\subsection{Enhanced quantum state sensor.} 
Figure~\ref{fig:fig1}\textbf{a} illustrates the architecture of the \textit{enhanced quantum state sensor} (EQSS), which integrates two complementary modules: (I) an optical bosonic reservoir that maps quantum inputs onto experimentally accessible observables through nonlinear evolution; and (II) an external neural-network readout that enhances generalisation and task adaptability. The synergy between two modules enables more efficient and universal analysis of quantum state properties compared with the standard quantum reservoir computing approach~\cite{Ghosh_2019,Swierczewski2025}. 

It should be noted that the robust generalisation is an essential requirement for advanced quantum machine learning applications, where even small variations in quantum states at the inference stage (and corresponding measured data) can have a decisive impact on performance. Meeting this requirement, we identify that the external neural network is the most appropriate solution to achieve high system performance.  

\textbf{\textit{Quantum module.}} In our theoretical framework, the reservoir is modelled as a lattice of coupled bosonic modes, confined within an optical microcavity and driven by a coherent external field. This configuration constitutes a hardware realisation of an optically driven quantum Bose–Hubbard system~\cite{Ciuti_2013, Carusotto_2013, Deuar_2021}. In the rotating frame of the driving laser, coherent evolution of the system is governed by the following dimensionless Hamiltonian
\begin{align}
\hat H
&= \sum_{j=1}^{N}\Bigg(
      \frac{U}{2}\,\hat b_j^{\dagger}\hat b_j^{\dagger}\hat b_j \hat b_j
      - \Delta_j \hat b_j^{\dagger}\hat b_j
      + F\,\hat b_j^{\dagger} + F^{*}\,\hat b_j
   \Bigg) \notag\\
 &\quad - \sum_{\langle i j \rangle} J_{ij}\big(\hat b_i^{\dagger}\hat b_j + \hat b_j^{\dagger}\hat b_i\big),
\label{eq:HR}
\end{align}
where $U$ denotes the strength of the on-site Kerr-type two-body interaction, $\hat{b}_j$ is the bosonic annihilation operator for reservoir site $j$, and $\Delta_j = \omega - \omega_j$ represents the detuning between the driving frequency $\omega$ and the natural mode frequency $\omega_j$. The total number of nodes is denoted by $N$, while $J_{ij}$ characterises the pairwise coupling between connected sites $i$ and $j$ with the connection denoted as $\langle i j \rangle$. The term $F$ represents the complex amplitude of the coherent driving field, uniformly distributed across all sites.

While this study focuses on an optical microcavity implementation, the underlying concept remains general and can be extended to other quantum platforms supporting higher-order interactions, including cross-Kerr terms ($\propto [b_i^\dagger \hat b_i \hat b_j^\dagger \hat b_j]$)~\cite{Kounalakis_2018} or parametric processes ($\propto[\hat b_i^2 + \hat b_i^{\dagger 2}]$)~\cite{DiCandia2023}. At this point, to maintain generality, no specific assumptions are imposed on the detuning distribution or the symmetry of the coupling matrix. Details of the parameters used during reservoir simulations are provided in the Methods Section (\textit{\nameref{sec:reservoir}}).

\textbf{\textit{Sensing protocol.}} The sensing protocol begins with reservoir initialisation (reservoir is initialised in vacuum state), followed by coherent laser driving $F$. At the same time, the photon field is damped, with the rate $\gamma$, as a result of coupling to the environment. The interplay of driving and dissipation stabilises the reservoir into a nonequilibrium steady state $\hat\rho_{ss}$, as shown in Fig.~\ref{fig:fig1}\textbf{b}. 

The target quantum states are subsequently injected into the reservoir in a step pulse-like, cascaded fashion~\cite{Lopez_2016}.  This injection perturbs the steady state and generates nonlinear temporal dynamics across the bosonic lattice. Assuming unidirectional cascaded coupling, the source state dissipates irreversibly into the reservoir without back-action, transferring correlations of the input state’s density matrix $\hat\rho_{in}$ into the reservoir node dynamics.  These correlations are mapped to measurable observables through continuous monitoring of average node occupation. Such optical signals can then be transduced into temporal digital or analogue signals via ultrafast photodetectors. It should be noted that sub-picosecond jitter and single-photon sensitivity are already technologically accessible with commercial streak cameras and superconducting nanowires~\cite{Korzh_2020}.

\textbf{\textit{Classical module.}}
\label{sec:classical} The central innovation introduced in this work, relative to conventional QRC, lies in the post-processing of the reservoir’s temporal responses by an auxiliary neural network module, as shown in Fig.~\ref{fig:fig1}\textbf{c}. As a classical module, a feed-forward neural network (FFNN) was employed for the post-processing. 
The FFNN can be expressed compactly as
\begin{equation}
f(\mathbf{x}) = \big(\phi_L \circ A_L \circ \cdots \circ \phi_1 \circ A_1\big)(\mathbf{x}),
\end{equation}
where $\mathbf{x} \in \mathbb{R}^{n}$ denotes the input vector of dimension $n$. 
For each layer $\ell = 1, \ldots, L$, the weight matrix and bias vector are given by 
$\mathbf{W}^{(\ell)} \in \mathbb{R}^{n_\ell \times n_{\ell-1}}$ and 
$\mathbf{b}^{(\ell)} \in \mathbb{R}^{n_\ell}$, respectively. 
The affine transformation for layer $\ell$ is defined as 
$A_\ell(\mathbf{y}) = \mathbf{W}^{(\ell)}\mathbf{y} + \mathbf{b}^{(\ell)}$, 
followed by an element-wise nonlinear activation 
$\phi_\ell : \mathbb{R}^{n_\ell} \to \mathbb{R}^{n_\ell}$. 
The composition operator ``$\circ$'' denotes the successive application of functions, $(g \circ f)(x) = g(f(x))$, representing the network as a hierarchical sequence of affine mappings 
and nonlinear transformations, enabling progressively abstract feature extraction \cite{dawid_2025}.

The input layer of FFNN receives an \(n\)-dimensional feature vector derived from the reservoir output. To construct the input vector, the non-perturbed reservoir dynamics $\langle \hat{n}_{ref,i}(t) \rangle$ is first subtracted from each node’s response, isolating the perturbation induced by the injected quantum state.  We would like to note that, unlike previous work ~\cite{Ghosh_2019}, the proposed extraction method requires an additional experiment on unperturbed reservoir dynamics to obtain the proper signal reference.

The analysis is then restricted to a time window in which the perturbed
dynamics are most pronounced. The background-corrected signal equal to $\tilde{n}_i(t)=\langle \hat{n}_i(t) \rangle-\langle \hat{n}_{ref,i}(t) \rangle$ is averaged over time intervals (bins) of duration~$\Delta t$. For the $k$-th bin, spanning in the range $[t_k,\, t_k + \Delta t)$, the averaged response is calculated as
$ \tilde{n}_i^{(k)} = \frac{1}{m_k} \sum_{t_j \in [t_k,\, t_{k+1})} \tilde{n}_i(t_j)$,
where $m_k$ denotes the number of discrete time samples, and $t_{k+1}=t_{k} + \Delta t$. 

Finally, the averaged responses from all nodes (and corresponding bins) are concatenated to form a single feature vector, $\mathbf{n} = \big[ \tilde{n}_1^{(1)}, \ldots, \tilde{n}_1^{(K)}, \tilde{n}_2^{(1)}, \ldots, \tilde{n}_N^{(K)} \big]^{\top}$. This vector is subsequently used as the classical module input ($\mathbf{x}=\mathbf{n}$), providing a compact, discrete representation of the reservoir state.

The FFNN considered in this work comprises $L$ hidden layers with $M_l$ neurons, where index $l$ indicates the hidden layer number, and an output layer of size determined by the target task. Through supervised training, this network learns to extract task-relevant features from the reservoir outputs, enabling adaptable operation across multiple tasks, illustrated schematically in Fig.~\ref{fig:fig1}\textbf{d}. Such architectures can be implemented directly on-chip, for instance, using field-programmable gate arrays (FPGAs) or application-specific integrated circuits (ASICs) \cite{Farsa_2025,electronics_9122193}. We found that a two-hidden-layer feed-forward neural network is the most universal and sufficient choice for enhanced sensing, considered in this work. More specialised or dedicated neural architectures, such as deep or convolutional networks, may be explored in the future to address increasingly complex quantum tasks.

As will be demonstrated in the following sections, the proposed hybrid quantum–classical architecture achieves significantly higher predictive accuracy than the standard bosonic QRC, while requiring substantially fewer physical resources, most notably reduced nonlinearity and a smaller reservoir size. To validate these findings, we performed numerical simulations of the considered system, using the positive-$\mathcal{P}$ formalism~\cite{Drummond_1980, Deuar_2002, Deuar_2006,Gardiner_Zoller_2010, Swierczewski2025}. 
This representation reformulates quantum evolution in terms of phase-space trajectories, 
enabling asymptotically exact simulation of driven-dissipative reservoir dynamics, perturbed by an incident quantum state. In addition, the positive-$\mathcal{P}$ formalism allows for reaching higher occupation regimes and for considering more modes than the standard master equation approach.

\subsection{Quantum reservoir modelling.}
The positive-$\mathcal{P}$ representation defines quantum dynamics as a stochastic evolution in a doubled complex phase space $(\boldsymbol{\alpha},\boldsymbol{\tilde{\alpha}}^{*})$, where the density matrix is expressed as $\hat{\rho}=\!\int d^{2N}\boldsymbol{\alpha}\,d^{2N}\boldsymbol{\tilde{\alpha}}\,
\mathcal{P}(\boldsymbol{\alpha},\boldsymbol{\tilde{\alpha}}^{*})
\hat{\Lambda}(\boldsymbol{\alpha},\boldsymbol{\tilde{\alpha}}^{*}),$ while the kernel $\hat{\Lambda}=\!\bigotimes_{j}\ket{\alpha_j}_j\bra{\tilde{\alpha}_j}_j/
\braket{\tilde{\alpha}_j}{\alpha_j}$ and $\mathcal{P}(\alpha, \alpha^\ast)$ is a real, positive probability density over stochastic configurations ~\cite{Drummond_1980,Deuar_2021}. 

For a driven–dissipative bosonic lattice, described by Hamiltonian (\ref{eq:HR}) coupled to an input source,
the quantum trajectory dynamics in the positive-$\mathcal{P}$ representation is given by
\begin{align}
\frac{d\alpha_i}{dt} &=A_{i}\alpha_i
- \text{i}U\alpha_i^2\tilde{\alpha}_i^*
   - \text{i}F  + \text{i}\!\sum_k J_{ik}\alpha_{k}+ \notag \\
   &\sqrt{-\text{i}U}\,\alpha_i\,\xi_i(t)
   - \sqrt{\gamma_s\gamma_i f(t)}\,{W}_i^{{in}} s, \notag\\[3pt]
\frac{d\tilde{\alpha}_i}{dt} &=A_{i}\alpha_i - \text{i}U\tilde{\alpha}_i^2\alpha_i^*
   - \text{i}F  + \text{i}\!\sum_k J_{ik}\tilde{\alpha}_{k} + \notag \\ &\sqrt{-\text{i}U}\,\tilde{\alpha}_i\,\tilde{\xi}_i(t)
   - \sqrt{\gamma_s\gamma_i f(t)}\,{W}_i^{{in}}\tilde{s},
\label{eq:alpha_dynamics}
\end{align}
where $A_{ik}=\text{i}\Delta_i- \frac{\gamma_i}{2}$ collect linear in $\alpha_{i}$, deterministic terms, while $\xi_i(t)$ and $\tilde{\xi}_i(t)$ are independent and real white-noise terms satisfying 
$\langle\xi_i(t)\xi_j(t')\rangle_S
 = \langle\tilde{\xi}_i(t)\tilde{\xi}_j(t')\rangle_S
 = \delta_{ij}\delta(t-t')$, $\langle\xi_i(t)\tilde{\xi}_j(t')\rangle_S = 0$ where $S$ denotes the system configurations taken into the stochastic average $\langle \cdot\rangle_S$ \cite{Deuar_2021}.
Here, $f(t)$ is a time-dependent coupling envelope (modelled as a rectangular step function),
$\text{W}_i^{\text{in}}\!\in[0,1]$ are random input weights.
The source mode evolves according to $\frac{ds}{dt} = -\,f(t)\frac{\gamma_s\eta}{2}s$ and $\frac{d\tilde{s}}{dt} = -\,f(t)\frac{\gamma_s\eta}{2}\tilde{s}$ \cite{Swierczewski2025}.  The parameter $\eta=\sum_i(\text{W}_i^{\text{in}})^2$ quantifies the total enhancement of dissipation, according to the input state information distribution across the reservoir nodes, where $\gamma_s$ denotes the rate of intrinsic source mode decay. 

The direct derivation of the considered equations for quantum neuromorphic computing systems is presented in Świerczewski et al.\cite{Swierczewski2025}. According to the Feynman-Kac theorem, quantum observables can be computed within the positive-$\mathcal{P}$ framework as ensemble averages, yielding exact values in the limit of large numbers of trajectories, taken respectively for the sites and sources from the vectors $\vec{v}_\alpha=\{\boldsymbol{\alpha},\boldsymbol{\tilde{\alpha}}^{*}\}$ and $\vec{v}_s=\{\boldsymbol{s},\boldsymbol{\tilde{s}}^{*}\}$. Therefore, the reservoir node occupation is defined as 
$\langle \hat{n}_i\rangle(t)=\Re\langle(\alpha_{i}(t)\tilde{\alpha}_{i}^{*}(t))\rangle_S$, averaged over stochastic configurations.
\\
\subsection{Quantum sensing with a hybrid quantum-classical sensor.}
\begin{figure*}[bht!]
    \centering
    \includegraphics[width=1\linewidth]{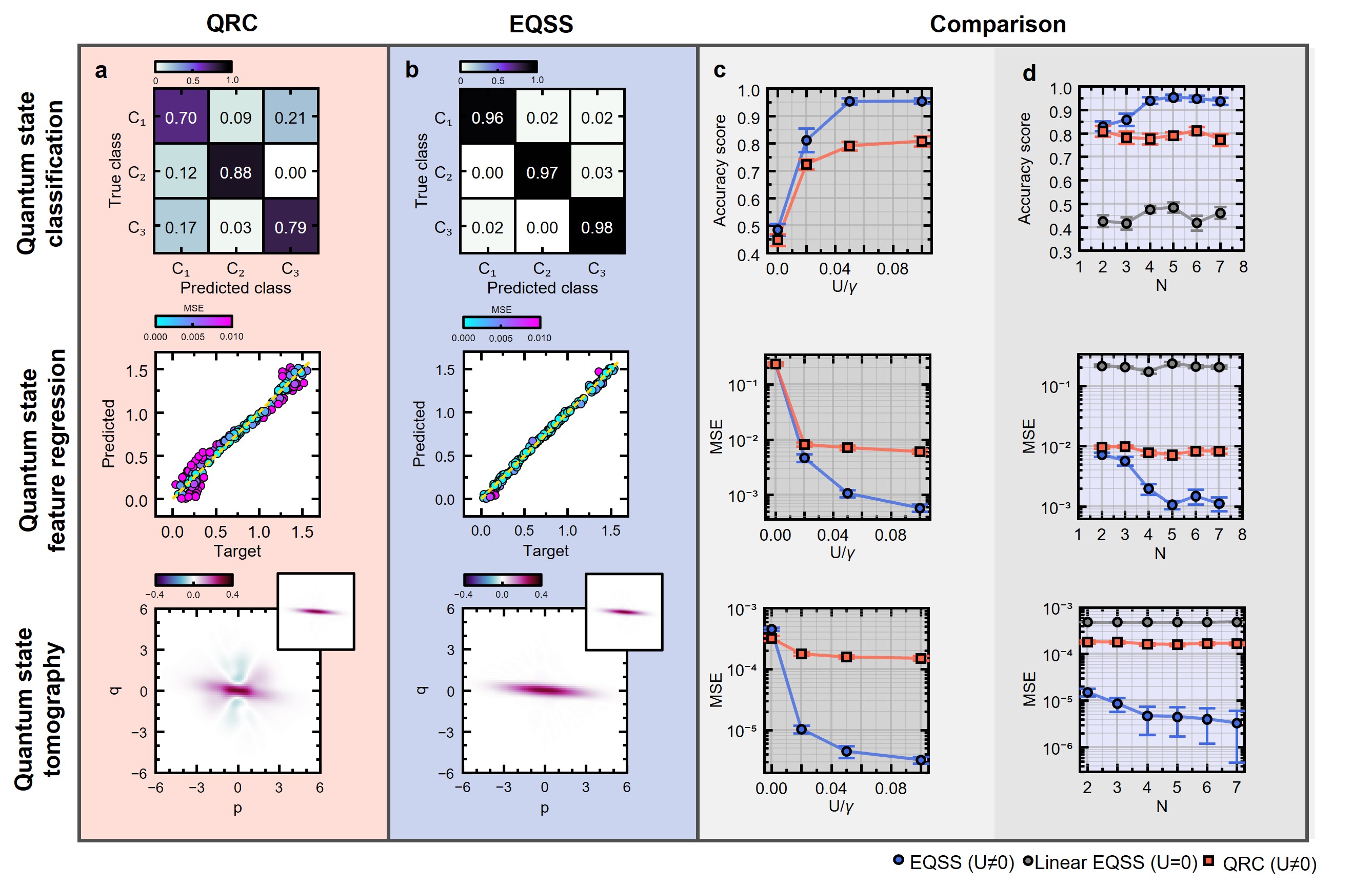}
\caption{\textbf{ Enhanced optical reservoir network as a multitask quantum state sensor.} 
\textbf{a}, Results of quantum state classification, single- and multi-parameter feature regression, and quantum state tomography obtained using the standard QRC approach. 
\textbf{b},~Corresponding results achieved with the EQSS detection strategy. Each panel in (\textbf{b}) directly corresponds to its counterpart in (\textbf{a}), sharing identical reservoir parameters and datasets. 
\textbf{c,d}, Dependence of classification accuracy and mean-squared error on reservoir nonlinearity $U$ and size $N$ for each benchmark task (columns correspond to specific tasks). 
Blue circles and orange squares represent results from EQSS and QRC, respectively, while grey circles indicate EQSS performance with a linear reservoir ($U=0$). 
Each data point was averaged over 10 independent training–testing realisations, with error bars denoting the standard deviation. The last row insets in panels \textbf{a} and \textbf{b}, represent the target state for quantum tomography task.
}
    \label{fig:fig2}
\end{figure*}

This section contains a systematic comparison between the standard QRC and the EQSS architectures across key benchmark tasks, including quantum state classification, single- and multi-parameter feature regression, and quantum state tomography, with the main results summarised in Fig.~\ref{fig:fig2}.

\textbf{\textit{Quantum states datasets.}}  
 In this work, we focus on the states that are particularly relevant to quantum metrology and information processing. The employed datasets were constructed from \textit{coherent}, \textit{squeezed}, and \textit{Schrödinger cat} states (750 samples in total for training and testing). The number of states belonging to the individual class across the whole dataset was set equal. States were defined as follows:

(I) The coherent state $\ket{\beta}$, representing the quantum state that most closely resembled a classical electromagnetic field was introduced as the eigenstate of the bosonic annihilation operator, satisfying $\hat{b}\ket{\beta} = \beta\ket{\beta}$, where $\beta = |\beta|e^{\mathrm{i}\varphi}$; 

(II) The squeezed vacuum state was expressed as
$\ket{\zeta} = \hat{S}(\zeta)\ket{0}$,
where the complex squeezing parameter $\zeta = r e^{2\mathrm{i}\theta}$ characterised both the magnitude $r$ and phase $\theta$ of squeezing. The corresponding squeezing operator was defined as
$\hat{S}(\zeta) = \exp\left[\frac{1}{2}\zeta^*\hat{b}^2 - \frac{1}{2}\zeta(\hat{b}^\dagger)^2\right]$;

(III) The Schrödinger cat state was represented as a coherent superposition of two coherent states with opposite phases,
 $\ket{\mathrm{cat}} = \mathcal{N}\!\left(\ket{\beta} + \ket{-\beta}\right)$,
where the normalisation factor was equal to
$\mathcal{N} = [2(1 + e^{-2|\beta|^2})]^{-1/2}$. where the complex coherent state amplitude is $\beta = |\beta|^{\text{i}\varphi}$.

Within each class, quantum states were generated with varying internal parameters (such as $\beta, \theta, \varphi, r$) to assess the sensor’s ability to generalise across intra-class variations. The states parameters were varied within the following intervals: for Schrödinger-cat states, $|\beta| \in (1.12, 1.38)$ and $\varphi \in \left(0, \frac{\pi}{2}\right)$; for squeezed states, $r \in (0.9, 1.1)$ and $\theta \in \left(0, \frac{\pi}{2}\right)$; and for coherent states, $|\beta| \in (1.03, 1.34)$ and $\varphi \in \left(0, \frac{\pi}{2}\right)$.   This particular range of parameter values for quantum states in the dataset was selected to achieve quantum states with mean photon densities spanning the same ranges, enabling correlation-based learning of quantum-state features. Further details, including the $\mathcal{P}$-representation of the considered quantum states, were provided in the work by Swierczewski et al. \cite{Swierczewski2025}.

\textbf{\textit{Quantum state classification.}}  
The first benchmark task was designed to evaluate whether the EQSS can discriminate between different types of quantum states better than QRC system (with a linear output layer). In the considered task,  squeezed (sq.) and Schrödinger cat (cat) states, with parameters within the following ranges $(|\beta| - |\beta_0| )^{2} + (\varphi - \varphi_0 )^{2} < R_{\mathrm{cat}}^{2}$
and $\quad
(r -r_0 )^{2} + (\theta - \theta_0 )^{2} < R_{\mathrm{sq}}^{2}$, forms a single class denoted as  $C_1$. States outside the specified parameter ranges were grouped into classes $C_2$ and $C_3$ based on their type (sq. or cat). This configuration reflects a practical experimental scenario in which identifying both the category and the characteristic features of an input quantum state is essential. Parameters $|\beta_0|$, $\varphi_0$ and $R_{\mathrm{cat}}$, were set to $2.35$, $1.15$ and $1.2$, while parameters $r_0$, $\theta_0$ and  $R_{\mathrm{sq}}$ were selected as $0.05$, $0.7$ and $1$. Here, the state parameters were selected to maximise linear inseparability among the considered classes.
We found that the proposed classification problem poses a challenge for a linear readout in a quantum reservoir, making it a particularly suitable choice for demonstrating the performance gains achievable through the hybrid quantum–classical approach.

The baseline performance of the system, obtained using the standard QRC approach without nonlinear post-processing, was evaluated for a five-node bosonic reservoir governed by Hamiltonian~(\ref{eq:HR}) with an interaction strength of $U/\gamma = 0.05$. The resulting confusion matrix, shown in Fig.~\ref{fig:fig2}\textbf{a}, indicates an overall accuracy of approximately $0.784$, with diagonal elements corresponding to correctly identified states. In this configuration, a linear readout layer was trained using multinomial logistic regression with an optimal $l_2$  regularisation rate. 
In this task, 250 distinct quantum states were used for training and testing (1:1  training-to-testing
ratio split). 

For EQSS, a nonlinear post-processing stage was introduced as a fully connected feed-forward neural network with one hidden layer containing $300$ GELU (Gaussian Error Linear Unit, $M_1=300$) neurons and a softmax output activation with three outputs. The weight optimisation was performed through minimising the categorical cross-entropy loss function. During the training stage, the reservoir observables were projected into probabilistic class assignments. To achieve high accuracy, the Adam optimiser with a learning rate of $dl=0.004$ and a minor weight decay of $dw=0.01$ was employed.

Hybrid configuration improves performance by capturing higher-order correlations and enhancing the separability of quantum states within the given class.  For the same datasets, EQSS achieves an accuracy exceeding $0.968$, as shown in Figs.~\ref{fig:fig2}\textbf{b}–\textbf{c}. The pronounced diagonal dominance in the confusion matrix indicates improved discrimination and generalisation across all classes. Figure~\ref{fig:fig2}\textbf{c} and \textbf{d} provide a quantitative comparison between QRC and EQSS, for varying values of nonlinear interaction $U$ and reservoir size $N$ respectively, averaged over ten independent runs. Error bars represent the standard deviation (obtained during the learning process with different weight initialisations), confirming the statistical robustness of the observed enhancement. The dependencies presented on panels \textbf{c} and \textbf{d} will be analysed in detail in a subsequent section.

\textbf{\textit{Quantum state feature regression.}}  
The second benchmark examined the capability of the EQSS framework to infer continuous parameters characterising incident optical quantum states. In contrast to classification tasks that assign discrete labels, regression aims to reconstruct continuous quantities that reflect intrinsic quantum-state properties. 

The dataset used in this task comprises only squeezed states (as described in the previous section). The measurement protocol was identical to that employed in the classification task, while the supervised learning stage was reformulated as a regression problem. The output layer dimension was adjusted to predict the squeezing phase of the squeezed states (single-output).
For this task, the single-hidden-layer network contains 250 GELU neurons. Comparing with the previous task, the number of neurons was reduced to avoid overfitting. Neural network was trained trough the minimization of the mean squared error, defined as $\mathrm{MSE} = \frac{1}{m}\sum_{i=1}^{m} |\theta_i - \tilde{\theta}_i|^2$, where $m$ denotes the number of validation states, and tilde notation is the predicted value of phase. During this task, the training-to-testing ratio was set to 3:10 (i.e., the number of test samples was higher than the number of training samples to avoid overfitting). The learning rate was 0.004, with a weight decay of 0.01. 

The corresponding results were summarised in middle panels of Figs.~\ref{fig:fig2}\textbf{a} and \textbf{b}, revealing that the hybrid approach achieved near-ideal correspondence between predictions and targets (yellow line). Averaged over ten independent runs (Fig.~\ref{fig:fig2}\textbf{c} and \textbf{d}), the EQSS consistently outperformed the standard QRC, demonstrating superior prediction (an order of magnitude lower MSE) in a continuous parameter regression task.\\

\textbf{\textit{Quantum state tomography.}}
The final benchmark focuses on reconstructing the Wigner function of an incident quantum state from the measured reservoir response. The training dataset consisted of various classes of quantum states, including squeezed, coherent, and Schrödinger-cat states, making this task particularly demanding due to the high generalisation requirements. The tomography task is the most challenging among the evaluated tasks, as it involves mapping reservoir dynamics onto the phase-space representation of the quantum state. Therefore, in this task, we use a deep neural network architecture with six hidden layers, containing $100$ (first four layers), $200$ and $64$ GELU neurons. 

To perform this task, the loss function is formulated to minimise the discrepancy between the target Wigner function, represented by an $(M \times M)$ matrix, and the neural network’s readout. For this purpose, either the mean squared error or the Huber loss function can be employed, depending on the desired robustness to outliers. 
The training-to-testing ratio was 14:1.

Bottom row of Figs.~\ref{fig:fig2}\textbf{a}–\textbf{b} present the reconstruction of a representative squeezed state, with the network trained using multiclass dataset containing squeezed coherent and cat states. As a representative example, the standard linear QRC approach yields a visibly distorted Wigner distribution (i.e., negativity that would not appear in the Wigner function of the squeezed quantum state), resulting in a high  MSE (Fig.~\ref{fig:fig2}\textbf{a}). In contrast, the enhanced hybrid architecture achieves near-ideal reconstruction with substantially reduced error, as shown in Fig.~\ref{fig:fig2}\textbf{b}, (an almost ideal reconstruction, as illustrated in the insets of tomography results as well as in Fig.~\ref{fig:fig2}\textbf{c}, showing the corresponding MSE values for $U/\gamma = 0.05$ and $N=5$), demonstrating the effectiveness of nonlinear post-processing in capturing quantum-state features.

In this context, we note that while this study focuses on the Wigner representation, the same methodology can readily be extended to other quasi-probability distributions, such as the Glauber–Sudarshan $\mathcal{P}$ or Husimi $\mathcal{Q}$ functions. 

\subsection{Reservoir size and degree of nonlinearity}
A central challenge in implementing QRC within optical platforms arises from the intrinsically weak optical nonlinearities and the limited scalability associated with enlarging the network of bosonic nodes that form the reservoir. To assess how the proposed hybrid architecture mitigates these constraints, a comparative analysis across varying nonlinear strengths and reservoir sizes was performed for the two analysed approaches.

Panels \textbf{c} and \textbf{d} of Fig.~\ref{fig:fig2} summarise the validation metrics, including classification accuracy and mean-squared error for regression and tomography, across the sequence of benchmark tasks introduced earlier.

As shown in Fig.~\ref{fig:fig2}\textbf{c}, even reservoirs with weak nonlinearities, e.g., $U/\gamma = 0.02$, achieve superior accuracy when coupled with the classical post-processing layer, outperforming standard QRC systems that employ nonlinearities five times stronger (see panel corresponding to the classification task). This improvement was consistently observed across all benchmarking tasks analysed. Additionally, it was verified by averaging over 10 independent realisations (with different training- and test-set selections), with error bars representing the associated standard deviation. Nevertheless, intrinsic nonlinearity remains indispensable. Reservoirs devoid of nonlinear interactions ($U = 0$) fail to match the performance of their nonlinear counterparts, even when augmented with external FFNN processing. This finding underscores the crucial role of the quantum reservoir in performing nonlinear transformations within the high-dimensional Hilbert space.

A similar trend is evident in Fig.~\ref{fig:fig2}(d), where the hybrid architecture sustains high performance with substantially fewer bosonic nodes. This result demonstrates that classical enhancement can effectively compensate for limited reservoir size, thereby improving scalability without compromising accuracy. Therefore, the obtained results highlight the potential of hybrid quantum–classical architectures to deliver high-performance quantum sensing under experimentally realistic hardware constraints.

\section*{\normalsize Discussion}

The results of this study establish the enhanced quantum state sensor as a robust and universal platform for next-generation quantum state detection. The proposed framework overcomes key bottlenecks that have limited the performance of the bosonic photonic quantum reservoir by integrating a bosonic quantum reservoir with a nonlinear classical readout. Across all investigated benchmarks, including quantum state classification, parameter regression, and continuous-variable tomography, the EQSS consistently outperforms standard QRC implementations, demonstrating higher accuracy, lower mean-squared error, and improved generalisation performance.

A key finding of this study is that classical post-processing can substantially enhance the computational expressivity of the quantum reservoir, particularly in regimes characterised by weak optical nonlinearities or limited reservoir size (see Fig.~\ref{fig:fig2}\textbf{c} and \textbf{d}). This result is of practical importance, as the strength of available nonlinear interactions in photonic platforms remains constrained by material properties. 
Nevertheless, intrinsic nonlinearity within the quantum reservoir remains essential for capturing higher-order correlations in the input states, highlighting the complementary rather than substitutive roles of the quantum and classical components.

In summary, the hybrid quantum–classical paradigm introduced here provides a scalable and experimentally viable pathway toward intelligent quantum photonic devices. By relaxing stringent material and designing requirements while preserving quantum nonlinearity, the EQSS bridges the gap between theoretical advances in quantum reservoir computing and their practical deployment in real-world technologies. Our approach provides a scalable alternative to variational quantum sensors and next-generation photonic architectures. Additionally, the demonstrated framework lays the groundwork for experimental implementation on existing integrated photonic platforms, enabling the development of adaptive, learning-based quantum sensors.
In view of recent milestone experiments based on integrated quantum circuits, which have already demonstrated the emergence of \textit{quantum advantage} across diverse computational tasks~\cite{McClean_2022, Riste_2017, Arute_2019}, providing the universal and fast quantum state detectors for further supporting technological progress can open new perspectives for the development of novel quantum technologies. 

\section*{\normalsize Methods}
\subsection{Quantum reservoir}\label{sec:reservoir}
We model the reservoir as a quantum Bose-Hubbard system defined on a two-dimensional square lattice, as referred in Equation (\ref{eq:HR}). The coupling between reservoir nodes is described by random nearest-neighbour Hermitian hopping terms, with amplitudes $J_{ij} = J^*_{ji}$. These hopping amplitudes are drawn from a uniform distribution in the range $J_{ij} \in (-1,1)$, and subsequently normalised by the spectral radius, i.e., the largest modulus of the system’s eigenvalues~\cite{Ghosh_2019}. On-site detunings, denoted by $\Delta_i$, are independently sampled from a uniform distribution $\Delta_i \in (0, 0.1\gamma)$. The corresponding system can be realised using a lattice of interconnected Kerr-quantum resonators or exciton-polariton modes in a lattice (or a disorder potential), as well as a system of interconnected nonlinear waveguides operating in the quantum regime.

\subsection{Quantum state sampling} \label{sec:sampling}
Within the phase-space formalism, rather than working directly with the density matrix $\hat{\rho}$ in the Fock state basis, a quantum state can be represented by a (quasi)probability distribution over a complex, doubled phase space $(\alpha,\tilde{\alpha}^*)$. Since the positive-$\mathcal{P}$ representation is not unique, identifying the most compact and efficient distribution for a given quantum state is generally nontrivial, however it can be simple for certain states, such as coherent states. For such states, only one pair $(\alpha,\tilde{\alpha})$ is needed to fully characterise the density matrix. More complex quantum states, such as squeezed vacuum states and optical Schrödinger cat states, require many pairs $(\alpha,\tilde{\alpha})$ to represent their density matrix, and these pairs are sampled from the probability distribution corresponding to the given state. In fact, any quantum state density matrix can be expressed in the \textit{canonical} form of the positive-$\mathcal{P}$ distribution \cite{PhysRevA.43.1153}, which is closely related to the Husimi-$\mathcal{Q}$ distribution and is guaranteed to be real and positive.
\\
\\

The positive-$\mathcal{P}$ distribution for a squeezed vacuum state,
\begin{equation}
\ket{\zeta}=\hat{S}(\zeta)\ket{0}, \text{ where } \hat{S}(\zeta)=\exp\!\left[\frac{1}{2}\zeta^*\hat{b}^2-\frac{1}{2}\zeta(\hat{b}^\dagger)^2\right],
\end{equation}

where the complex squeezing parameter is $\zeta=re^{2\mathrm{i}\theta}$. Using the definition of the canonical positive-$\mathcal{P}$ distribution and following Olsen \& Bradley \cite{Olsen_2009}, one finds
\begin{equation}
P_{\mathrm{sq,can}}(\alpha,\tilde{\alpha}^*)=
\frac{e^{-\nu_x^2/(e^{-r}\cosh r)}}{\sqrt{\pi e^{-r}\cosh r}}
\frac{e^{-\nu_y^2/(e^{r}\cosh r)}}{\sqrt{\pi e^{r}\cosh r}}
\frac{e^{-|\delta|^2}}{\pi}.
\end{equation}

This distribution can be sampled by defining
\[
\alpha = e^{\mathrm{i}\theta}\nu+\delta, \qquad
\tilde{\alpha} = e^{\mathrm{i}\theta}\nu-\delta,
\]
where
\[
\delta=\frac{1}{\sqrt{2}}(q_1+\mathrm{i}q_2), \qquad
\nu=\nu_x+\mathrm{i}\nu_y
=\sqrt{\frac{e^{-r}\cosh r}{2}}\,q_3
+\mathrm{i}\sqrt{\frac{e^{r}\cosh r}{2}}\,q_4.
\]
Here, $q_1,q_2,q_3,q_4$ are independent Gaussian random variables with zero mean and unit variance. Unlike the coherent state case, accurate estimates of observables such as the mean density or quadratures for a squeezed vacuum state require averaging over many stochastic samples.

The Schrödinger cat state is defined as a coherent superposition of two coherent states with opposite phases,
\[
\ket{\mathrm{cat}}=\mathcal{N}\bigl(\ket{\beta}+e^{\mathrm{i}\theta}\ket{-\beta}\bigr),
\]
with normalization factor
\[
\mathcal{N}=\frac{1}{\sqrt{2\left(1+e^{-2|\beta|^2}\cos\theta\right)}}.
\]

The density matrix $\hat{\rho}_{\mathrm{cat}}$ can be expressed in the coherent-state basis as a sum of four terms proportional to
$\ket{\beta}\bra{\beta}$,
$\ket{\beta}\bra{-\beta}$,
$\ket{-\beta}\bra{\beta}$,
and
$\ket{-\beta}\bra{-\beta}$.
To evaluate the canonical positive-$\mathcal{P}$ distribution at a given phase-space point $(\alpha,\tilde{\alpha}^*)$, one computes the scalar products
\[
\bra{\tfrac{1}{2}(\alpha+\tilde{\alpha})}\ket{\beta},
\qquad
\bra{\tfrac{1}{2}(\alpha+\tilde{\alpha})}\ket{-\beta},
\]
and combines them with the appropriate coefficients. For an arbitrary cat state, the resulting canonical distribution is
\begin{align}
\begin{split}
P_{\mathrm{cat,can}}(\alpha,\tilde{\alpha}^*)
&=|\mathcal{N}|^2
e^{-\frac{1}{2}\left(|\alpha|^2+|\tilde{\alpha}|^2\right)}
e^{-|\beta|^2}
\\
&\quad\times
\left[
e^{\frac{(\alpha+\tilde{\alpha})^*}{2}\beta}
+e^{\mathrm{i}\theta}
e^{-\frac{(\alpha+\tilde{\alpha})^*}{2}\beta}
\right]^2.
\end{split}
\label{p_can_cat}
\end{align}

This distribution can be sampled to generate stochastic realizations of $\alpha$ and $\tilde{\alpha}^*$, thereby enabling the construction and numerical simulation of the Schrödinger cat state within the positive-$\mathcal{P}$ framework.

\subsection{Numerical simulations of the reservoir dynamics} For the numerical treatment of the  positive-$\mathcal{P}$ equations governing the reservoir dynamics (see Eq. (\ref{eq:alpha_dynamics})), we have chosen the semi-implicit midpoint algorithm \cite{DRUMMOND1983211}. The algorithm's applicability for solving stochastic differential equations in the positive-$\mathcal{P}$ framework has been studied in \cite{Deuar2021multitime}, where a detailed description of the method and its applicability has been provided. The simulation time step was set to $\tau = 0.05$ and the simulation window was $T = 25$.  

\clearpage{}
\newpage{}

\section*{\normalsize References}

\vspace{15mm}

\bibliographystyle{naturemag}
\bibliography{bib}


\end{multicols}

\scriptsize
\section*{Data availability}
\noindent All data that supports the conclusions of this study are included in the article. The data presented in this study are available from the corresponding author upon reasonable request.

\section*{Acknowledgments} 
\noindent   This work was supported by the National Science Center, Poland, under the following projects, 
This work was supported by Quantum Optical Networks based on Exciton-polaritons - (Q-ONE) funding from the HORIZON-EIC-2022-PATHFINDER CHALLENGES EU programme under grant agreement No. 101115575. AO acknowledges support from the National Science Center, Poland, project No. 2024/52/C/ST3/00324.


\section*{Competing interests} 
\noindent The authors declare no competing interests.

\section*{Additional information} 
\noindent{\bf Correspondence and requests for materials} should be addressed to A.O.

\end{spacing}
\end{document}